\documentclass{sig-alternate-05-2015}

\usepackage{balance}  
\usepackage{graphics} 
\usepackage[T1]{fontenc}
\usepackage{txfonts}
\usepackage{mathptmx}
\usepackage[pdftex]{hyperref}
\usepackage{color,soul}
\usepackage{booktabs}
\usepackage{textcomp}
\usepackage{topcapt}

\usepackage{microtype} 
\usepackage{tabularx}

\setcopyright{rightsretained}
\conferenceinfo{WSDM'16,}{February 22--25, 2016, San Francisco, CA, USA.}
\isbn{978-1-4503-3716-8/16/02.}
\doi{http://dx.doi.org/XXXX.XXXX}

\begin{document}

\title{Static Ranking of Scholarly Papers using Article-Level Eigenfactor (ALEF)}

\numberofauthors{3}
\author{%
  \alignauthor{Ian Wesley-Smith\\
    \affaddr{The Information School \\University of Washington}\\
    \affaddr{Seattle, WA 98195 USA}\\
    \email{iwsmith@uw.edu}}\\
  \alignauthor{Carl T. Bergstrom\\
    \affaddr{Department of Biology \\University of Washington}\\
    \affaddr{Seattle, WA 98195 USA}\\
    \email{cbergst@uw.edu}}\\
  \alignauthor{Jevin D. West\\
    \affaddr{The Information School \\University of Washington}\\
    \affaddr{Seattle, WA 98195 USA}\\
    \email{jevinw@uw.edu}}\\
}

\maketitle
 
\begin{abstract}
Microsoft Research hosted the \href{https://wsdmcupchallenge.azurewebsites.net/}{2016 WSDM Cup Challenge} based on the Microsoft Academic Graph.
The goal was to provide static rankings for the articles that make up the graph, with the rankings to be evaluated against those of human judges.
While the Microsoft Academic Graph provided metadata about many aspects of each scholarly document, we focused more narrowly on citation data and used this contest as an opportunity to test the Article Level Eigenfactor (ALEF), a novel citation-based ranking algorithm, and evaluate its performance against competing algorithms that drew upon multiple facets of the data from a large, real world dataset (122M papers and 757M citations).
Our final submission to this contest was scored at 0.676, earning second place.

\end{abstract}



\section{Introduction}
The scholarly literature comprises at least 50 million articles, and some estimates place this tally
as high as 150 million.
Today, scientific advances occur at ever increasing rates, technology 
enables a wealth of new approaches to investigation, and scholars publish their work in a rapidly diversifying variety of venues. 
Despite all this, scholars continue to search for papers as if it were 1960.
Manually tracing citation chains remains an important mode of search and discovery, with researchers often choosing which papers to read based off little more than the title and names of the authors.
The fundamental problem was succinctly described by Vannevar Bush
70 years ago:
``The summation of human experience is being expanded at a prodigious rate, and the means we use for threading through the consequent maze to the momentarily important item is the same as was used in the days of square-rigged ships.'' \cite{bush1945atlantic}
Transportation has improved immensely since 1945, our information overload problem has only worsened.

Over the past few decades, massive increases in digital storage and computational power have opened the door to automated systems for helping scholars find important material in the vast academic literature. Such systems aim to identify documents that are both relevant to a user's interests and of sufficiently high importance to merit attention. Determining relevance is a problem of matching; determining importance is a problem of ranking. Once documents are assembled into a vast network by hyperlinks among them, the connectivity of the linked documents provides extensive contextual information that can be used for ranking. Google's PageRank algorithm \cite{PageEtAl98,Brin} provides a perfect example. Before Google, most search engines looked for relevant search results without considering the importance of the pages they returned. Finding the most important websites required human-curated web directories such as that offered by Yahoo at the time. Google showed us that latent within the hyperlink graph of the world wide web lies the information needed to solve both the problems of determining relevance and of determining importance. 

The scholarly literature has its own graph structure, with papers as nodes and links laid down by the practice of scholarly citation. We can use this structure to automate the process of scholarly discovery \cite{bergstrom2010use}. We have spent several years applying network-based methods to citation graphs to determine the importance of entities in those graphs \cite{RosvallAndBergstrom08,RosvallAndBergstrom10,RosvallAndBergstrom11,WestEtAl10,WestEtAl13,RosvallEtAl14}. Although our methods were originally developed to assess the importance of journals, we have recently developed a modified ranking methodology which operates on individual articles: Article-Level Eigenfactor (ALEF) \cite{west2016ALEF}.

The 2016 WSDM Cup Challenge presented an excellent opportunity to evaluate and extend our methods, while comparing their performance to other approaches.
This is the first time we have evaluated ALEF in a static ranking challenge.
We found that ALEF was effective by itself, more so than simpler network methods.
We also developed a basic method of extending our rankings to authors. This provided additional score coverage and a slightly higher score in the competition.

\section{Related Work}
Researchers have been using the structure of citation networks to gather information about scholarly articles for many decades. 
One of the first attempts at a similarity metric was \emph{bibliographic coupling}, which measured the similarity between two documents based on the number of papers cited by both of them \cite{Kessler1963}.
This approach, though novel, had some shortcomings.
The most substantial of these limitations derives from the time-directed nature of citation graphs.
Since a paper can only cite papers published before it, an early paper will be missing many of the citations included in a later paper on the same subject.
As a result, bibliometric coupling matches documents that were written at roughly the same time; the similarity metric fails to evolve with the field. 
To address these shortcomings,  Small \cite{Small1973} reversed the directionality of citation tracing to introduce a method known as \emph{co-citation similarity}. The co-citation approach determines document similarity based on the frequency with which two documents are cited together in other papers.
Unlike bibliographic coupling, co-citation similarity evolves and changes as fields change, providing a similarity measure that is based on the current academic milieu, rather than one from twenty years ago. 

Although researchers initially used bibliographies to calculate similarity metrics, they soon had the idea of measuring citation flows between journals to understand importance. The core notion in this area, introduced by Bonacich \cite{Bonacich1972}, is {\em eigenvector centrality}. Bonacich wanted to find the important nodes in a social network; his idea was that important nodes are those which are linked to by other important nodes. While this may sound circular, it is merely recursive and can be computed by straightforward matrix operations. Bonacich's approach found its greatest value in algorithms for searching the world wide web: \emph{HITS} by Kleinberg \cite{Kleinberg1999} and \emph{PageRank} by Brin and Page \cite{Brin}. 
Eigenvector centrality algorithms afford two equivalent interpretations. We can view them as voting algorithms in which each node votes iteratively for the nodes to which it links, or as tracking algorithms in which we follow the trajectory of a random walker on the network and compute the steady-state distribution of the walker's position. Under the latter interpretation, the time spent at each node reflects its relative importance, thereby generating rankings for all nodes. We will use the random walker interpretation here.

Long before eigenvector centrality approaches were used for ranking websites, researchers explored their utility for ranking scholarly journals when applied to citation networks. Early papers included work by Pinski and Narin \cite{Pinski1976} and Liebowitz and Palmer \cite{Liebowitz1984}. A substantial subsequent literature refined these initial approaches \cite{
Kalaitzidakis2003,PalaciosHuerta2004,Kodrzycki2006,Walker2006,Bollen2006}. Our own work developing the evaluationr algorithm followed in this tradition \cite{Bergstrom07,WestEtAl10}. Eigenvector centrality methods in general, and PageRank analogues in particular, can be applied directly to journal-level citation networks. Article-level citation networks are more challenging because they are directed acyclic graphs in which following a random walker leads inexorably back in time toward the earliest papers. Nonetheless, several author have explored this approach \cite{Chen2007,Gori2007,Vellino2009}. 
The actual efficacy of these attempts is difficult to measure. However, a qualitative analysis found that when adjusting the algorithm to deal with the directed acyclic nature of the network, an eigenvector centrality approach could identify ''scientific gems'', papers that may not be cited often but are of great importance in their field \cite{Chen2007}. Here we present the latest iteration of this process. ALEF has its conceptual foundations in eigenvector centrality approaches, but deals with the problem of directed acyclic graphs in a new way. 

\section{Contest Description}
The impetus for this work was the \href{https://wsdmcupchallenge.azurewebsites.net/}{2016 WSDM Cup Challenge}, hosted by Microsoft Research. 
The challenge is to statically rank articles in
the Microsoft Academic Graph \cite{Sinha2015}. The August 20\textsuperscript{th}, 2015 version of the graph, which we used throughout, consists of over 122 million papers, 123 million authors and 757 million citations. It provides metadata including article title, authors, publication date, journal or conference name, author affiliations, citation data, and the URL from which the paper had been retrieved. Teams were to assign a value from 0 to 1 for each paper, with higher scores corresponding to more important papers.
Rankings were evaluated by comparing the pair-wise ranking for a set of papers that have been ranked by human experts.  
If the algorithm concurred with the experts on the relative importance of two papers, then the algorithm was considered to have correctly ranked this pair.
An algorithm's performance was expressed as the fraction of the time that the algorithm agreed with the human experts in a pairwise comparison of papers. A score of .676 means means that the algorithm agreed with the human judges 67.6\% of the time.

The contest was divided into two phases. This paper largely reports on the results from Phase I; Phase II is ongoing. In Phase I, the expert data was randomly split into two data sets: an evaluation (training) set that teams could query repeatedly to develop their algorithms, and a test set that was withheld from teams and used to rank the final submissions. 
Neither the evaluation set nor the test set  were directly available to the contestants. Contestants could only observe the performance of their rankings as submitted to Microsoft Research.

\section{Algorithm}
Our aim in this study was to determine whether a simple, low-parameter algorithm that used only one facet of the metadata---the citation network---could more effectively and robustly rank scholarly papers than more complicated approaches.

Our approach is first to calculate a citation-based score for each paper using ALEF.
We then use these article scores to generate scores for each author.
We then combine the scores of all the authors of a paper to generate an author score for that paper.
Finally, we combine the citation and author scores to generate a final score for each paper.

\subsection{Author Level Eigenfactor}
Citations have long been considered one of the best sources of information about a paper's influence.
Although there is substantial criticism of using raw citation counts as a measure of importance,
 methods derived from \emph{PageRank} and \emph{HITS} alleviate some of this concern.
We used a method developed at Ume{\aa} University and the University of Washington called {\em Article-Level Eigenfactor} or (ALEF) \cite{west2016ALEF}.

Like \emph{PageRank}, ALEF simulates a random walk on the citation network. Citation networks are directed, (nearly) acyclic graphs: a paper can only cite papers written before it. Thus a random walker that follows the citation trail will inevitably move to earlier and earlier papers as it progresses, and will spend the majority of its time at older citations. Because citation and web link matrices may be non-irreducible or nearly so, most algorithms allow the random walker to ``teleport'' to a new location in the network with some frequency, 15\% of the time in the original \emph{PageRank} implementation. In directed acyclic graphs, this further serves to help the random walker escape dead ends and moves walker forward in time to more recent papers.

ALEF ameliorates this problem of moving backward in time by shortening the number of steps the random walker takes between teleportation steps, and by modifying how the teleportation steps are made. For this version of ALEF, the random walker takes only one step before teleporting again.  We tested other variants of this parameter, but found that additional steps on the citation graph did not improve the ranking. The ALEF teleporation process has the random walker teleport to links rather than nodes. 

First, we teleport to a random link in the network. 
Next, instead of beginning the walk, we move with equal probability either ``upstream'' of the link to a node that cites this one, or ``downstream'' of the link to a node cited by this one.
This is effectively an undirected walk at this step, allowing us to move either back or forwards in time.
We then resume our time-directed walk, scoring nodes based on how often we arrive at them.
We believe that this approach helps to alleviate the problem of unduly weighting older papers.

\subsection{Author Score}
ALEF is only able to assign non-zero scores to papers that have been cited and are well integrated into the larger citation network. This is particularly problematic for the most recently released papers that have not yet had time to accumulate citations. In an attempt to increase the number of papers that we can rank, we supplement the ALEF score for each paper with author scores derived from the ALEF scores of the authors. 

We create two types of author scores: an individual scholar's author score (IA) and a paper's author score (PA). The IA score is the mean non-zero ALEF score of all the papers for a single author, while the PA score is the mean non-zero IA score for all the authors of that paper. We use means rather than medians so that a big hit paper has a substantial impact on a scholar's IA score, and a big name coauthor has a substantial impact on a paper's PA score.

\subsection{Combining Scores}
Using ALEF scores alone yielded a performance of $0.693$, while author scores alone yielded a performance of $0.655$. The final step in computing a paper's score is computing a weighted average of these components. To determine the optimal weights we performed a manual parameter sweep. We found that 70:30, 50:50, and 30:70 combinations of ALEF and author scores yields performances of 0.699,0.693, and 0.691 respectively. Thus
in our final submission we used 70\% citation score and 30\% author score.

\subsection{Randomizing unranked papers}
Since the contest scoring is performed by comparing the scores of two papers, a pair of papers that both have zero-valued scores will necessarily fail to match the ordering chosen by the humans experts. 
Even a random guess should be correct half the time, and {\em if} having the wrong order is no worse than having a tie, such guessing should improve the scoring overall. Since our method only scored 54\% of papers even with author scores added, we assigned small random scores for all unranked papers. This is similar to regularization methods or adding noise to inject "serendipity" into the results.  
To ensure these randomized scores did not outscore a paper we had an actual ranking for, we found the minimum non-zero score of ranked papers and to all zero-valued papers we randomly assigned scores drawn from a uniform distribution on $[0, \mbox{minval} * 0.999]$. This had no effect on the performance of our rankings, and we removed this randomization step from our Phase II submission. We mention it here for completeness.

\subsection{Code \& Environment}

One of the strengths of ALEF is its computational efficiency.  The calculation is relatively fast and the mechanics are easy to explain. Subtracting the read/write times, the scores can be calculated in about 2.5 hrs on a standard 2.4 GHz machine. 
Thus our scoring system could be run on a daily basis to keep up with a growing corpus. We have written our code in C++ (ALEF step) and Python (author-level step).  The Python code uses the NumPy and SciPy\cite{SciPy} packages.  However, our method does consume significant memory when calculating on the a graph as big as the MAS graph (approximately 50 GB). The code is freely available at github.com/jevinw/.

\begin{figure}[htbp]
	\centering	 
	\includegraphics[width=\columnwidth]{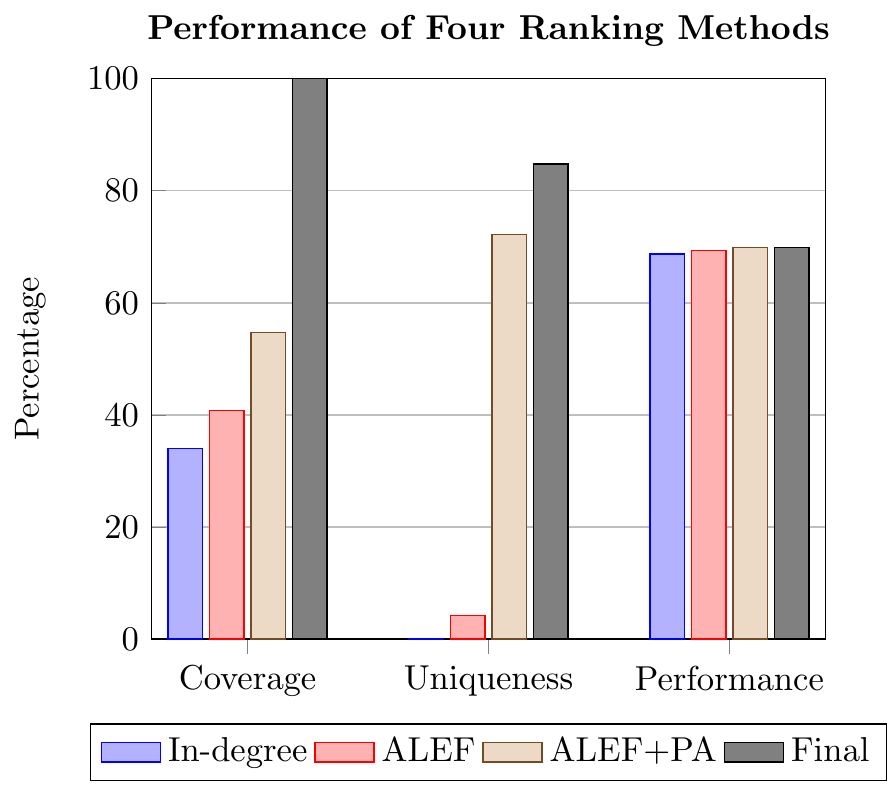}
    \caption{Statistics of four result sets described in the text.
    Coverage indicates the percentage of papers to which we were able to assign scores. Uniqueness indicates the percentage of uniquely valued scores in our rankings. Performance is the percentage of comparisons in which our rankings agreed with those of human experts in the evaluation set.}
    \label{fig:results}
\end{figure}

\section{Results}		

Figure~\ref{fig:results} provides a summary of evaluation set performance for 4 (out of a total of 45) ranking methods that we evaluated in Phase I of the competition. For each method, we report: coverage, the percentage of the corpus for which we could provide non-zero scores; 
uniqueness, the percentage of scores that were uniquely valued; and performance, the percentage of correct comparisons when our rankings were judged against those of human experts.  

The first method shown is {\em In-degree}: the raw number of citations  number of citations that a paper receives, divided by the total number of citations in the dataset to obtain a value on $[0,1]$. The second is our basic {\em ALEF} score; with this approach we were able to provide rankings for 40\% of the papers and we received a judged score of 0.693.
The third is {\em ALEF+PA}: here we took a weighted avergae of each paper's ALEF scores and its paper author score (PA), using 70\% ALEF and 30\% PA when both were available and 100\% of whatever we had when only one was available. 
This improved our coverage, providing an additional 17,180,522 scores (+34.3\%) and increasing our judged score to 0.699.
Finally, we generated random scores for un-scored papers, which is why our final submission has 100\% coverage. This did not improve our judged score.

Our score on the test dataset was 0.676, putting us in second place after Phase I of the competition.

\section{Discussion}

One of the big lessons from the competition was the utility of simple, low-parameter methods. During the evaluation stage of Phase I, it was unclear whether such methods would be effective; for much of this period our percentage of correct comparisons hovered just below 70\% and we barely remained among the top 20 teams. When the competition shifted from the evaluation stage to the test stage, our success percentage dropped from 69.9,\% to 67.6\%, but our rank position rose from 15 to 2. This suggests that many of the other approaches may have been overfitting during evaluation stage. Switching evaluation sets was a useful feature of the competition in that it to encouraged more robust rankings and discouraged overfitting. 

Since the evaluation and test data sets were not revealed, we do not know which papers were ranked by human experts nor even how many papers were ranked in this way. Nor do we yet know the specifics of the methods used by the other winning teams. Consequently, it is impossible at this point to know why certain methods performed better than others. We will learn more when the top teams publish their methods; additionally there would be a benefit to releasing these two labeled sets of data so that we could better interpret the results of the competition. 

One of the main goals of this work was to evaluate ALEF's ability to rank papers.
When we compared ALEF to degree centrality, we found that it did score higher, though not by much: .693 v .687. 
One area where ALEF performed substantially better than degree centrality is in coverage, in that this method generates 20\% more scores due to the dynamics of teleportation. 
Although the score increase is slight, when combined with the substantially larger coverage this implies that ALEF is more useful for static ranking, as it can rank a greater number of documents with slightly higher accuracy than degree centrality. The Spearman's correlation between degree centrality and ALEF is $\rho=0.97$, but these seemingly minor differences in net score can be extremely important for relative ranking \cite{west2010big}. 

Inferring author scores from citation data appeared useful, providing a .006 increase in our evaluation score and allowed us to rank an additional 14\% of the corpus. 

A list of the top papers, as ranked by ALEF, is initially alarming.
The first two entries are the same paper (Protein Measurement With the Folin Phenol Reagent, by Lowry et al 1951), the third is a journal ({\em N Engl J Med}), and the fourth is the text "In the Press". 
Looking closely, however, we find that these are each listed as papers in the underlying graph database. Thus our approach is generating the correct scores, given the underlying graph data. 
To validate this, we looked up the top paper using other citation databases and found that it is a seminal methods paper on protein measurement. For example, Google Scholar reports a mind boggling 200,000 citations. Finding it at the top of any citation based measure of impact is unsurprising; the dual appearance results from duplicated entries in the original data.
The third entry is not a paper at all, but rather the New England Journal of Medicine, a highly respected and influential journal.
Though the URL associated with this paper is a \href{http://www.researchgate.net/publication/259762349_N_Engl_J_Med}{ResearchGate entry}, it is uclear how entry was constructed and how it was assigned 132,239 citations (while only citing 4 documents).
The fourth entry, \emph{In the press}, has similar characteristics to the third.

Finally, the judging method created a potentially interesting insight.
A pair-wise ranking evaluation means a completely random ranking should result in a score of approximately 50\%.
We tested this idea by submitting a result set that consisted of random scores drawn uniformly from $[0, 1]$, which was scored at 0.526.
If we consider that the first place entry received a score of 0.683, this means the best ranking in the competition is only 30\% better than random chance.
Interpreting this fact is a bit difficult.
On one hand, only being 30\% better than random chance sounds quite bad.
However, an A-B relative ranking evaluation metric is as favorable as possible for a random ranking, because it doesn't penalize ridiculous guesses and more than close misses. Thus the seemingly lackluster upgrade rate of 30\% may not be as bad as it first appears.
We are hopeful that Phase II, which will use the rankings as part of Microsoft Academic Search and use usage data to validate the results, will provide more insight into the performance of various algorithms.

\section{Conclusion}
This contest allowed us to determine several useful things.
First, ALEF performed well, especially when the evaluation datasets were varied. This suggests that a large part of the signal in the Microsoft Academic Graph lies in the citation network alone. 
Given the substantial success of machine learning methods in other domains, it is interesting to see that network based methods  perform well on scholarly papers, at least for static rankings. 
However, one should not discount machine learning techniques as there is no reason to believe they are inherently untenable in this domain. Rather, we may simply not understand how to apply them appropriately yet. Furthermore, machine learning and network based methods are not mutually exclusive; presumably skillful integration of both would give even better results than seen here.

Second, the idea of inferring other scores, such as author scores, from citation scores provides additional coverage and a slight improvement in performance.  

Finally, this contest pushed the limits of our current ALEF implementation. This dataset, with a citation graph of 38 million nodes and over 600 million edges, was the largest we have processed.
As datasets improve and more articles are published and added to the graph, we will need to improve our implementation to keep pace so that rankings can be updated in real time.  

In sum, we found the contest to be a useful exercise and hope to see further competitions in the future.

\section{Acknowledgements}
This work was supported by the Metaknowledge Network funded by the John Templeton Foundation. We would also like to thank Martin Rosvall for his extensive role in developing the original ALEF algorithm.

\bibliographystyle{abbrv}
\bibliography{wsdm_2016}

\begin{thebibliography}{10}

\bibitem{Bergstrom07}
C.~Bergstrom.
\newblock Measuring the value and prestige of scholarly journals.
\newblock {\em College \& Research Libraries News}, 68(5):314--316, 2007.

\bibitem{bergstrom2010use}
C.~T. Bergstrom.
\newblock Use ranking to help search.
\newblock {\em Nature}, 465:870, 2010.

\bibitem{Bollen2006}
J.~Bollen, M.~A. Rodriquez, and H.~Van~de Sompel.
\newblock Journal status.
\newblock {\em Scientometrics}, 69(3):669--687, 2006.

\bibitem{Bonacich1972}
P.~Bonacich.
\newblock Factoring and weighting approaches to clique identification.
\newblock {\em Journal of Mathematical Sociology}, 2:113--120, 1972.

\bibitem{Brin}
S.~Brin and L.~Page.
\newblock The anatomy of a large-scale hypertextual web search engine.
\newblock {\em Computer networks}, 56(18):3825--3833, 2012.

\bibitem{bush1945atlantic}
V.~Bush and A.~W.~M. Think.
\newblock The atlantic monthly.
\newblock {\em As we may think}, 176(1):101--108, 1945.

\bibitem{Chen2007}
P.~Chen, H.~Xie, S.~Maslov, and S.~Redner.
\newblock {Finding scientific gems with Google's PageRank algorithm}.
\newblock {\em Journal of Informetrics}, 1(1):8--15, jan 2007.

\bibitem{Gori2007}
M.~Gori and A.~Pucci.
\newblock {Research paper recommender systems: A random-walk based approach}.
\newblock {\em Proceedings - 2006 IEEE/WIC/ACM International Conference on Web
  Intelligence (WI 2006 Main Conference Proceedings), WI'06}, pages 778--781,
  2007.

\bibitem{SciPy}
E.~Jones, T.~Oliphant, P.~Peterson, et~al.
\newblock {SciPy}: Open source scientific tools for {Python}, 2001--.

\bibitem{Kalaitzidakis2003}
P.~Kalaitzidakis, T.~Stegnos, and T.~P. Mamuneas.
\newblock Rankings of academic journals and institutions in economics.
\newblock {\em Journal of the European Economic Association}, 1:1346--1366,
  2003.

\bibitem{Kessler1963}
M.~M. Kessler.
\newblock {Bibliographic Coupling Between Scientific Papers}.
\newblock {\em American Documentation (pre-1986)}, 14(1):10, 1963.

\bibitem{Kleinberg1999}
J.~M. Kleinberg.
\newblock {Authoritative Sources in a Hyperlinked Environment}.
\newblock 46(5):604--632, 1999.

\bibitem{Kodrzycki2006}
Y.~K. Kodrzycki and P.~D. Yu.
\newblock New approaches to ranking economics journals.
\newblock {\em B.E. Journal of Economic Analysis and Policy}, 5(1):Article 24,
  2006.

\bibitem{Liebowitz1984}
S.~J. Liebowitz and J.~P. Palmer.
\newblock Assessing the relative impacts of economics journals.
\newblock {\em Journal of Economic Literature}, 22:77--88, 1984.

\bibitem{PageEtAl98}
L.~Page, S.~Brin, R.~Motwani, and T.~Winograd.
\newblock The {PageRank} citation ranking: {B}ringing order to the web.
\newblock {\em WWW7 / Computer Networks}, 30:107--117, 1998.

\bibitem{PalaciosHuerta2004}
I.~Palacios-Huerta and O.~Volij.
\newblock The measurement of intellectual influence.
\newblock {\em Econometrica}, 72:963--977, 2004.

\bibitem{Pinski1976}
G.~Pinski and F.~Narin.
\newblock {Citation influence for journal aggregates of scientific
  publications: Theory, with application to the literature of physics}.
\newblock {\em Information Processing {\&} Management}, 12(5):297--312, 1976.

\bibitem{RosvallAndBergstrom08}
M.~Rosvall and C.~T. Bergstrom.
\newblock Maps of random walks on complex networks reveal community structure.
\newblock {\em Proceedings of the National Academy of Sciences, USA},
  105:1118--1123, 2008.

\bibitem{RosvallAndBergstrom10}
M.~Rosvall and C.~T. Bergstrom.
\newblock Mapping change in large networks.
\newblock {\em PLoS One}, 5:e8694, 2010.

\bibitem{RosvallAndBergstrom11}
M.~Rosvall and C.~T. Bergstrom.
\newblock Multilevel compression of random walks on networks reveals
  hierarchical organization in large integrated systems.
\newblock {\em PLoS One}, 6:e18209, 2011.

\bibitem{RosvallEtAl14}
M.~Rosvall, A.~V. Esquivel, A.~Lancichinetti, J.~D. West, and R.~Lambiotte.
\newblock Memory in network flows and its effects on spreading dynamics and
  community detection.
\newblock {\em Nature communications}, 5, 2014.

\bibitem{Sinha2015}
A.~Sinha, Z.~Shen, Y.~Song, H.~Ma, D.~Eide, B.-j.~P. Hsu, and K.~Wang.
\newblock {An Overview of Microsoft Academic Service (MAS) and Applications}.
\newblock In {\em Proceedings of the 24th International Conference on World
  Wide Web Companion (WWW 2015 Companion)}, pages 243--246, 2015.

\bibitem{Small1973}
H.~Small.
\newblock {Co-Citation in Scientific Literature: A new measure of the
  relationship between two documents}.
\newblock {\em Journal of the American Society for Information Science},
  24(4):265--269, 1973.

\bibitem{Vellino2009}
A.~Vellino.
\newblock Recommending journal articles with pagerank ratings.
\newblock ACM, 2009.

\bibitem{Walker2006}
D.~Walker, H.~Xie, K.-K. Yan, and S.~Maslov.
\newblock Ranking scientific publications using a simple model of network
  traffic.
\newblock {\em Journal of Statistical Mechanics: Theory and Experiment},
  6:P06010, 2007.

\bibitem{west2010big}
J.~West, T.~Bergstrom, and C.~T. Bergstrom.
\newblock Big macs and eigenfactor scores: Don't let correlation coefficients
  fool you.
\newblock {\em Journal of the American Society for Information Science and
  Technology}, 61(9):1800--1807, 2010.

\bibitem{west2016ALEF}
J.~West, M.~Rosvall, D.~Vilhena, and C.~Bergstrom.
\newblock Ranking and mapping article-level citation networks.
\newblock {\em In Prep.}, 2016.

\bibitem{WestEtAl10}
J.~D. West, T.~C. Bergstrom, and C.~T. Bergstrom.
\newblock The eigenfactor metricstm: A network approach to assessing scholarly
  journals.
\newblock {\em College \& Research Libraries}, 71(3):236--244, 2010.

\bibitem{WestEtAl13}
J.~D. West, M.~C. Jensen, R.~J. Dandrea, G.~J. Gordon, and C.~T. Bergstrom.
\newblock Author-level eigenfactor metrics: Evaluating the influence of
  authors, institutions, and countries within the social science research
  network community.
\newblock {\em Journal of the American Society for Information Science and
  Technology}, 64(4):787--801, 2013.

\end{thebibliography}

\end{document}